\documentclass[11pt]{article}
\usepackage{amsmath}
\usepackage{graphicx}
\begin{document}
\title{Hidden-variable models for the spin singlet. \\
II. Local theories violating Bell and Leggett inequalities}
\author{Antonio Di Lorenzo\footnote{dilorenzo@infis.ufu.br}\\
Instituto de F\'{\i}sica, Universidade Federal de Uberl\^{a}ndia,\\
 38400-902 Uberl\^{a}ndia, Minas Gerais, Brazil}
\date{April 14, 2011}
\maketitle
\begin{abstract}
Three classes of local hidden-variable models that violate both Bell and Leggett inequalities are presented. 
The models, however, do not reproduce the quantum mechanical predictions, hence they are 
experimentally testable. It is concluded that on one hand neither Bell or Leggett inequality fully captures the essential 
counterintuitiveness of quantum  mechanics, while on the other hand the hypothesis of outcome independence and 
that of locality are uncorrelated. 
\end{abstract}
\maketitle
A hidden-variable model is a hypothetical theory where a system is specified by some parameters, the 
so-called ``hidden variables'', which substitute or complement the description of quantum theory. The latter 
specifies the state of a system through a wavefunction $\Psi$ (or more generally a density matrix $\rho$). 
From a historical point of view, we expect quantum mechanics to be eventually superseded by some more 
fundamental theory that reproduces the quantum mechanical predictions within the range of  
the current experimental precision (or even exactly) 
in the situations where quantum mechanics applies, which so far are practically  
all known situations. In order to be considered seriously, a hidden-variable theory 
must provide more than this. It should either make 
predictions disagreeing with the ones of quantum mechanics in some suitable situation or provide a measurement 
procedure for the additional variables.  

To the best of our knowledge, the only full fledged hidden-variable model, reproducing the quantum mechanical 
predictions by construction, is the one due to Bohm \cite{Bohm1952}, which is not local. 
All other hidden-variable models try more modestly to reproduce the quantum mechanics of the simplest system: 
spin 1/2 or polarization of light. 
While it is easy to reproduce the results for a single spin through a simple model \cite{Bell1964}, 
considering two entangled spins proves already a difficult problem, at least if one wants to retain locality. 
Bell \cite{Bell1964} proved the incompatibility of all deterministic local hidden-variable models 
with quantum mechanics. The proof was reelaborated in Ref.~\cite{Clauser1969} to make it experimentally testable.  
The incompatibility between quantum mechanics and the models complying with Bell's hypotheses was demonstrated 
through the formulation of an inequality, obeyed by the latter theories and violated by quantum mechanics. 
Bell inequality was shown to apply to a class of 
local stochastic model \cite{Bell1971,Clauser1974}, 
provided that the hypothesis of determinism is replaced by the so-called hypothesis of outcome-independence 
\cite{Jarrett1984,Shimony1990}. Recently, it was demonstrated that a stochastic model satisfies 
outcome-independence if and only if it is deterministic at some more fundamental level \cite{Hall2011}.
Experiments \cite{Bellexp} confirm the predictions of quantum mechanics, 
barring some unlikely experimental loophole.

In recent times, another inequality, called the Leggett inequality, was formulated \cite{Leggett2003}, 
which applies to a distinct class of local models: 
The hypothesis of outcome independence is dropped in favor of another hypothesis, namely that for fixed values of 
the hidden variables (assumed to be unit vectors $\mathbf{u}$ and $\mathbf{v}$), the outcome at a detector should obey Malus's law,
 i.e. the marginal probability of observing the outcome $\sigma$ when the detector is measuring the spin component 
$\mathbf{a}\cdot\mathbf{S}$, should be $P(\sigma|\mathbf{u},\mathbf{a})=(1+\sigma \mathbf{a}\cdot\mathbf{u})/2$. 
By considering the same class of models, Branciard and collaborators \cite{Branciard2008} 
arrived to a different inequality, which we call the Branciard inequality. 
In this case experiments \cite{Leggettexp} confirm the predictions of quantum mechanics as well. 

Both Bell and Leggett-Branciard inequalities exclude not only the local models they consider, but also all 
other models, local or non-local, characterized by some additional hidden parameters $\lambda'$ such that upon 
integration over these latter variables the marginal probabilities satisfy either Bell or Leggett hypotheses. 
This truism is usually not stressed in experiments violating Bell inequalities, but is often addressed as a disproof
of ``non-local realism'' when discussing  Leggett inequality, due to the fact that Ref.~\cite{Leggett2003} 
derived the inequality by assuming unnecessarily some special form of non-locality \cite{DiLorenzo2011b}. 

This is the second paper in a series of three. In the first paper \cite{DiLorenzo2011c}, a  non-local hidden variable model 
for the spin singlet was presented, and the model was compared to pre-existing non-local hidden-variable models 
\cite{Groblacher2007b,DeZela2008,Hall2010}. 
The non-locality was shown to result in the possibility of instantaneous communication, 
provided that the hidden parameters could be measured without being perturbed. In the third paper \cite{DiLorenzo2011e}, 
a class of local models reproducing the quantum mechanical predictions will be presented. 
The models in the first and third paper reproduce exactly 
the probability distribution predicted by quantum mechanics, thus the only way to make them experimentally testable would 
be to specify how to measure the hidden variables. In the present paper, we discuss a family of models which are local, 
violate Bell \cite{Bell1964,Clauser1969,Clauser1974}, Leggett \cite{Leggett2003}, 
and Branciard \cite{Branciard2008} inequalities, yet do not reproduce quantum mechanics, and thus are experimentally 
testable, even without specifying a measurement protocol for the additional parameters. 

The quantity experimentally accessible in experiments measuring the polarizations of two photons or the spin projection of 
two spin 1/2 particles is the joint probability $P(\sigma,\tau|\Psi,\Sigma)$. Here, $\Sigma=\{\mathbf{a},\mathbf{b}\}$ 
specifies the setup of the detectors, i.e. two unit vectors, and $\Psi$ the preparation of a singlet state. 
In order to keep the formulas short and to stress the functional dependence on the parameters, we write this probability as 
$P_{\sigma,\tau}(\Psi,\Sigma)$. 
Let us consider the following families of hidden-variable models for a spin singlet:  
When the detectors are set to 
measure the spin components along the directions $\mathbf{a}$ and $\mathbf{b}$, 
the joint conditional probability of observing outcomes $\sigma$ and $\tau$ for a bipartite system prepared in a spin singlet 
state and characterized by the hidden variables $\lambda$ (in two of the families 
$\lambda=\{\mathbf{u},\mathbf{v}\}$, two unit vectors)  is 
\begin{subequations}\label{eq:fund}
\begin{align}
\label{eq:fund1}
P^{F}_{\sigma,\tau}(\mathbf{u},\mathbf{v},\Psi,\Sigma)
=&\frac{1}{4}+\frac{
\eta\left[\sigma f(\mathbf{u}\cdot\mathbf{a})+\tau f(\mathbf{v}\cdot\mathbf{b})\right]
-\sigma\tau \mathbf{a}\cdot\mathbf{b} 
}{4(1+\eta)}, \\
\label{eq:fund2}
P^{S}_{\sigma,\tau}(\lambda,\Psi,\Sigma)
=&\frac{1}{4}-\sigma\tau \frac{
 \mathbf{a}\cdot\mathbf{b}+\left(\mathbf{a}\times\mathbf{b}\right)\cdot\mathbf{p}(\lambda) 
}{4\sqrt{1+p_m^2}},\\
\label{eq:fund3}
P^{T}_{\sigma,\tau}(\mathbf{u},\mathbf{v},\Psi,\Sigma)
=&\frac{1}{4}-\sigma\tau \frac{
 \mathbf{a}\cdot\mathbf{b}
+\zeta\left(\mathbf{a}\cdot\mathbf{u}\right)^3\left(\mathbf{b}\cdot\mathbf{v}\right)^3 }{4},
\end{align}
\end{subequations}
where (a) the function $f$ satisfies $|f(x)|\le 1/2$ for $|x|\le 1$ and is otherwise arbitrary, so that the probability is positive for all values of the parameters, with $\eta$ an arbitrary positive number and 
(b) $\mathbf{p}(\lambda)$ is a vector dependent on the hidden-variables $\lambda$ and $p_m^2=\mathop{sup}_{\lambda} |\mathbf{p}(\lambda)|^2$, while (c) $\zeta$ is a positive constant. 
For later convenience we define $\varepsilon=\eta/(1+\eta)$.  
We shall refer to the three families as first, second and third hidden-variable model (FHV,SHV,THV).  
The distribution $\rho(\lambda|\mathbf{a},\mathbf{b})$ of the hidden variables is required to be 
independent of the settings of the detectors, i.e., 
\begin{equation}\label{eq:locdistr}
\rho(\lambda|\mathbf{a},\mathbf{b})=\rho(\lambda).
\end{equation} 
Actually, the functions multiplying $\sigma$ and $\tau$ in Eq.~\eqref{eq:fund1} need not be equal. 
We chose them so just because of the symmetry of the setup, and for the same reason we consider 
$\rho(\mathbf{u},\mathbf{v})=\rho(\mathbf{v},\mathbf{u})$. However, these hypotheses will not be exploited at any point, except for THVs, where we take $\rho^{T}(\mathbf{u},\mathbf{v})=\delta(\mathbf{u}+\mathbf{v})/4\pi$. 
In order to reproduce the well tested experimental fact that the marginal probability of observing 
either outcome is 1/2, we require that 
\begin{equation}
\int d\mathbf{u}d\mathbf{v} \rho^{F}(\mathbf{u},\mathbf{v}) f(\mathbf{u}\cdot\mathbf{a})= 0 \ .
\end{equation}
The theories specified by Eqs.~\eqref{eq:fund} and \eqref{eq:locdistr} 
are local, since (i) the settings of the detectors do not influence the distribution of the hidden variables and (ii) 
from Eq.~\eqref{eq:fund} each of the marginal probabilities 
is independent of the setting of the remote detector: they are indeed  
\begin{equation}\label{eq:margprob1}
P^{F}_\sigma(\mathbf{u},\mathbf{v},\mathbf{a},\mathbf{b})=\frac{1+\varepsilon\sigma f(\mathbf{u}\cdot\mathbf{a})}{2}
\end{equation} 
and  $P^{S,T}_\sigma(\lambda,\mathbf{a},\mathbf{b})=1/2$. 
Furthermore, these families of hidden-variable models do not satisfy the hypothesis of outcome independence, since the conditional probabilities are 
\begin{align}
\label{eq:condprob1}
P^{F}_{\sigma|\tau}(\mathbf{u},\mathbf{v},\Psi,\Sigma)=&\frac{1}{2}\left\{1+
\sigma\frac{\eta f(\mathbf{u}\cdot\mathbf{a})-\tau \mathbf{a}\cdot\mathbf{b} 
}{1+\eta+\eta\tau f(\mathbf{v}\cdot\mathbf{b})}\right\}, \\
\label{eq:condprob2}
P^{S,T}_{\sigma|\tau}(\lambda,\Psi,\Sigma)=&2 P^{S,T}_{\sigma,\tau}(\lambda,\Psi,\Sigma).
\end{align}
Nor the hypothesis of compliance with Malus's law is verified, since by putting $\mathbf{u}=\mathbf{a}$ in 
Eq.~\eqref{eq:margprob1} we have 
\begin{equation}
\frac{1}{2}{\left[1+\varepsilon\sigma f(1)\right]}\neq \frac{1}{2}{\left[1+\sigma\right]},
\end{equation}
while, as stated, $P^{S,T}_\sigma(\lambda,\mathbf{a},\mathbf{b})=1/2$. 
Thus the models under consideration satisfy the locality hypothesis, required for both Bell and Leggett inequalities, 
but do not satisfy either of the additional requirements 
(outcome independence for the Bell inequality and Malus's law for the Leggett inequality) 
needed in order for the inequalities to be derived. 
This is the opposite case of the earlier paper \cite{DiLorenzo2011c} in this series, where locality was not satisfied but both outcome independence 
and Malus's law were. 
For brevity we refer to the models considered 
in deriving the Bell inequality, characterized by 
\begin{align}
\label{eq:fundbell}
&P_{\sigma,\tau}(\lambda,\Psi,\Sigma)=
\frac{\bigl[1+ P(\sigma|\lambda,\mathbf{a})\bigr]\bigl[1+ P(\tau|\lambda,\mathbf{b})\bigr]}{4}, 
\end{align}
as BHV, and to the models violating the Leggett-Branciard inequalities, 
which satisfy 
\begin{align}
\label{eq:fundlegg}
&P_{\sigma,\tau}(\lambda,\Psi,\Sigma)=
\frac{1+\sigma \mathbf{u}\cdot\mathbf{a}+\tau\mathbf{v}\cdot\mathbf{b}+ \sigma\tau C_{\mathbf{u},\mathbf{v}}(\mathbf{a},\mathbf{b})}{4},
\end{align}
as LHV. The distribution of the hidden variables is assumed local but otherwise arbitrary in all
these models, according to Eq.~\eqref{eq:locdistr}. 

For a spin singlet, quantum mechanics predicts the correlator
\begin{equation}
C^{QM}(\Psi,\Sigma)=-\mathbf{a}\cdot\mathbf{b} . 
\end{equation}
Our models predict instead 
\begin{subequations}
\label{eq:corr}
\begin{align}
\label{eq:corr1}
C^{F}(\Psi,\Sigma)=&-\frac{\mathbf{a}\cdot\mathbf{b}}{1+\eta} ,\\
\label{eq:corr2}
C^{S}(\Psi,\Sigma)=&
-\frac{\mathbf{a}\cdot\mathbf{b}+(\mathbf{a}\times\mathbf{b})\cdot\overline{\mathbf{p}}}{\sqrt{1+p_m^2}} ,\\
\label{eq:corr3}
C^{T}(\Psi,\Sigma)=&-\left(1-\frac{3\zeta}{35}\right)\mathbf{a}\cdot\mathbf{b}
+\frac{2\zeta}{35}(\mathbf{a}\cdot\mathbf{b})^3 ,
\end{align}
\end{subequations}
with $\overline{\mathbf{p}}\equiv \int d\lambda\, \rho^{S}(\lambda)\mathbf{p}(\lambda)$. 

Let us now proceed to demonstrate that all the models considered violate Bell, Leggett, and Branciard inequalities 
for suitable values of $\eta$. 
In the Clauser-Horne inequality (which is equivalent to Bell inequality), the following linear combination is considered 
\begin{align}\nonumber
\mathcal{E}_{\mathbf{a},\mathbf{b},\mathbf{a}',\mathbf{b}'}\equiv
\left|
C(\mathbf{a}\cdot\mathbf{b})\!+\!C(\mathbf{a}\cdot\mathbf{b}')\!+\!C(\mathbf{a}'\cdot\mathbf{b})\!-\!
C(\mathbf{a}'\cdot\mathbf{b}')
\right|.
\end{align}
BHVs satisfy $\mathcal{E}^{BHV}_{\mathbf{a},\mathbf{b},\mathbf{a}',\mathbf{b}'}\le 2$, while quantum mechanics provides 
$\mathcal{E}^{QM}=2\sqrt{2}$ for a proper choice of the orientations.
For FHVs, we have that 
$\mathcal{E}^{F}=\mathcal{E}^{QM}/(1+\eta)$, 
thus the Bell inequalities are violated by FHVs for $\eta\le \sqrt{2}-1\approx 0.4142$. 
Interestingly, for SHVs, the contributions 
of $\left(\mathbf{a}\times\mathbf{b}\right)\cdot\overline{\mathbf{p}}$ in Eq,~\eqref{eq:corr2} 
to the Bell-CH parameter $\mathcal{E}^{S}$ 
cancel out for this optimal configuration. It is then immediate to verify that Bell inequality is violated by SHVs as well, e.g., 
with $\mathbf{a},\mathbf{b},\mathbf{a}',\mathbf{b}'$ 
equal to the optimal quantum values, provided that $p_m^2< 1$. Finally, for THVs, 
still considering the optimal quantum values, 
one has $\mathcal{E}^{T}=2\sqrt{2}-\zeta/(3\sqrt{2})>2$ for $\zeta<12-6\sqrt{2}\approx 3.5417$.

On the other hand, the quantity appearing in Leggett inequality 
as formulated in Ref.~\cite{Leggett2003,Groblacher2007b} is 
\begin{align}
\mathcal{F}(\phi)\equiv\left|C_\mathbf{p}(\phi)+C_\mathbf{p}(0)\right|
+\left|C_{\mathbf{p}'}(\phi)+C_{\mathbf{p}'}(0)\right| ,
\end{align}
where $\mathbf{p}$ and $\mathbf{p}'$ represent orthogonal planes, and  
\[
C_\mathbf{p}(\phi)\equiv \int  d\mathbf{a}\,d\mathbf{b} \ 
\mu_{\mathbf{p},\phi}(\mathbf{a},\mathbf{b}) C(\mathbf{a},\mathbf{b}) ,
\]
with the normalized measure $\mu_{\mathbf{p},\phi}$ in the integral fixing $\mathbf{a}$ and $\mathbf{b}$ to have a relative angle 
$\phi$ and to lie in the plane $\mathbf{p}$. 
Leggett inequality gives $\mathcal{F}^{LHV}(\phi)\le 4-(4/\pi)\sin{|\phi/2|}$. 
Quantum mechanics predicts 
$\mathcal{F}^{QM}(\phi)= 2(1+\cos{\phi})$ and violates the inequality for $|\phi|\le2 \arcsin{(1/\pi)}$, 
while the maximum violation happens 
for $|\phi_m|=2 \arcsin{(1/2\pi)}$, when 
\begin{equation}
\mathcal{F}^{QM}(\phi_m)
-4+\frac{4}{\pi}\sin{\left|\frac{\phi_m}{2}\right|}=\frac{1}{\pi^2}\approx 0.1013.
\end{equation} 
FHVs predict a violation for all values of $\eta\le 2\pi^2-1-2\pi\sqrt{\pi^2-1}\approx 0.0267$, occurring when 
\begin{equation}
\left(\sin{\left|\frac{\phi}{2}\right|}-\frac{1+\eta}{2\pi}\right)^2\le \frac{(1+\eta)^2}{4\pi^2}-\eta .
\end{equation}
The maximum violation for fixed $\eta$ corresponds to 
$|\phi'_m|=2\arcsin{\frac{1+\eta}{2\pi}}$, when  
\begin{equation}
\mathcal{F}^{F}(\phi'_m)-4+\frac{4}{\pi}\sin{\left|\frac{\phi'_m}{2}\right|}=
\frac{-4\eta}{1+\eta}+\frac{1+\eta}{\pi^2}.
\end{equation}
When considering SHVs, by choosing $\mathbf{p}$ and $\mathbf{p}'$, respectively, parallel and orthogonal to $\overline{\mathbf{p}}$ 
in Leggett inequality, we have 
\begin{equation}
\mathcal{F}^{S}(\phi)=\frac{2(1+\cos{\phi})+|\overline{\mathbf{p}}|\sin{\phi}}{\sqrt{1+p_m^2}} . 
\end{equation}
Because of the additional term in the numerator, after considering $\eta=\sqrt{1+p_m^2}-1$ and 
substituting all values of $\phi$ found above for which the FHVs do not satisfy Leggett inequality, we have that 
the latter is violated \emph{a fortiori} by SHVs as well. 
The optimal violation, however, involves the solution of a third degree 
equation and will not be reported here. 
THVs violate Leggett inequality as well, e.g., 
for  $\phi=\phi_m$ and $\zeta<70\pi^4/(40\pi^6-18\pi^4+6\pi^2-1)\approx 0.1855$.

Finally, let us consider Branciard inequality \cite{Branciard2008}. 
The relevant parameter is 
\begin{equation}
\mathcal{G}(\phi) \equiv \frac{1}{3}\sum_{i=1}^3|C(\mathbf{a}_i,\mathbf{b}_i)+C(\mathbf{a}_i ,\mathbf{b}'_i)|,
\end{equation}
where $\mathbf{a}_i$ are three orthogonal vectors, $\mathbf{b}_i$ and $\mathbf{b}'_i$ 
lie in the plane formed by $\mathbf{a}_i$ and $\mathbf{a}_{i+1}$ (the indexes summing modulo three) and form an angle $\pm\phi/2$ with $\mathbf{a}_i$.  
In Ref.~\cite{Branciard2008} it is demonstrated that 
\begin{equation}
\mathcal{G}^{LHV}(\phi)\le 2-\frac{2}{3} \sin{\left|\frac{\phi}{2}\right|}.
\end{equation}
Quantum mechanics predicts $\mathcal{G}^{QM}(\phi)=2|\cos{(\phi/2)}|$, thus violates the inequality for 
\( \sin{\left|\phi/2\right|}\le 3/5\) ,
and the maximum violation occurs for 
$\sin{|\phi''_m/2|}=1/\sqrt{10}$, when 
\begin{equation}
\mathcal{G}^{QM}(\phi^{''}_m)-2+\frac{2}{3} \sin{\left|\frac{\phi''_m}{2}\right|}=\frac{2}{3}\sqrt{10}-2\approx 0.1082 . 
\end{equation}
For $\eta\le 3/(2\sqrt{2})-1\approx 0.0607$, FHVs violate the inequality for all \(\phi\) satisfying 
\begin{equation}
\left(\sin{\left|\frac{\phi}{2}\right|}-\frac{(1+\eta)^2}{3}\right)^2\le \frac{1-16\eta-8\eta^2}{9}\ .
\end{equation}
The maximum violation, for fixed $\eta$, occurs when 
$\sin{|\phi^{'''}_m/2|}=(1+\eta)/\sqrt{9+(1+\eta)^2}$,
\begin{equation}
\mathcal{G}^{F}(\phi^{'''}_m)-2+\frac{2}{3} \sin{\left|\frac{\phi^{'''}_m}{2}\right|}
=\frac{2}{3}\frac{\sqrt{9+(1+\eta)^2}}{1+\eta}-2.
\end{equation}
Interestingly, even though the degree of violation of Branciard inequality is somewhat 
greater than that of Leggett inequality, the latter applies to a wider range of $\eta$. 
SHVs violate Branciard inequality for the same values of the angle between \(\mathbf{a}\) and \(\mathbf{b}\) 
when identifying \(\eta=\sqrt{1+p_m^2}-1\), since the terms proportional to \(\mathbf{a}\times\mathbf{b}\)
in Eq.~\eqref{eq:corr2} cancel out in  \(\mathcal{G}\). THVs, finally, violate Branciard inequality in a range of 
\(\phi\) and \(\zeta\) which has a complicated expression and will not be related here. Suffice to say that 
for \(\phi_m^{''}\) the inequality is violated for \(\zeta<175(10-3\sqrt{10})/216\approx 0.4158\).  

We notice that after the integration over the hidden variables, the FHVs are indistinguishable from 
the quantum state where a singlet state is produced with probability \(1-\varepsilon\) 
and a zero spin incoherent mixed state with probabiliy \(\varepsilon\). 
This could be attributed to imperfect preparation or detection. 
FHVs will thus fit experimental data better than the quantum mechanical predictions for a pure singlet state. 
This platitude should not lure one to believe that FHVs 
are serious candidates to replace quantum mechanics. 
To start with, they apply only to entangled spins, while 
it is highly desirable to have a theory encompassing all possible predictions: spin, 
orbital angular momentum, position, momentum, etc., such as Bohm's \cite{Bohm1952}. 
Furthermore, the function $f(x)$ should be specified along with a coherent mathematical formalism 
that justifies the form of the probability in Eq.~\eqref{eq:fund}. Finally a protocol 
for measuring $\mathbf{u}$ and $\mathbf{v}$ should be provided, so that the form of 
$f(x)$ can be tested by experiments.  
SHVs, on the other hand, make a prediction that should be easily falsified by experiments: 
by switching the orientations of the detectors the correlator should change according to Eq.~\eqref{eq:corr2}. 
THVs predict a correlator [see Eq.~\eqref{eq:corr3}] 
with a functional dependence on the angle between \(\mathbf{a}\) and 
\(\mathbf{b}\) which is distinct from the simpler quantum mechanical expression for the spin singlet, thus they 
can be falsified as well.

In conclusion, we have provided three classes of local stochastic hidden-variable theories 
that violate the three inequalities of Bell, Leggett, and Branciard. 
In Ref.~\cite{DiLorenzo2011c} a class of non-local models which satisfy outcome-independence and violate all three inequalities was presented. 
Here, on the contrary, we presented a class of local models which do not obey 
outcome-independence. This proves conclusively that 
outcome-independence and locality are logically uncorrelated concepts. 
Furthermore, we have shown, by means of a counterexample and not by abstract reasoning, 
that the violation of the three inequalities does not falsificate the so-called hypotheses of 
``local realism'' or ``non-local realism''. Finally, the predictions of quantum mechanics are not exactly reproduced by our models. On one hand, this means that they can be experimentally tested; on the other hand this fact implies that Bell and Leggett-Branciard inequalities do not capture fully the features of quantum mechanics. 
A natural question is now whether there are local theories that reproduce exactly quantum mechanics. 
It will be answered in the affirmative in the last installment of this series \cite{DiLorenzo2011e}.

I acknowledge stimulating discussions with M. J. W. Hall. 
This work was supported by Funda\c{c}\~{a}o de Amparo \`{a} Pesquisa do 
Estado de Minas Gerais through Process No. APQ-02804-10.

\end{document}